\documentclass[aps,prb,twocolumn,floatfix,superscriptaddress,showpacs]{revtex4-1}

\usepackage{graphicx}
\usepackage{transparent}
\usepackage{xcolor}
\usepackage[centertags]{amsmath}
\usepackage{amsfonts}
\usepackage{amssymb}
\usepackage{bm}
\usepackage{dcolumn}
\usepackage{epstopdf}

\renewcommand{\vec}[1]{\boldsymbol{#1}}
\newcommand{\ang}{~\text{\AA}}
\newcommand{\kel}{~\text{K}}
\newcommand{\rhos}[1]{\rho_s^{\text{#1}}}

\begin{document}

\title{Local Superfluidity at the Nanoscale}

\author{B. Kulchytskyy}
\affiliation{Department of Physics, McGill University, Montreal, H3A
2T8, Canada}

\author{G. Gervais}
\affiliation{Department of Physics, McGill University, Montreal, H3A
2T8, Canada}

\author{A. \surname{Del Maestro}}
\affiliation{Department of Physics, University of Vermont, Burlington, VT 05405,
USA}

\date{\today}

\begin{abstract}
Motivated by the search for an experimentally realizable high density and
strongly interacting one dimensional quantum liquid, we have performed quantum
Monte Carlo simulations of bosonic helium-4 confined inside a nanopore with
cylindrical symmetry. By implementing two numerical estimators of superfluidity
corresponding to capillary flow and the rotating bucket experiment, we have
simultaneously measured the finite size and temperature superfluid response of
${}^4$He to the longitudinal and rotational motion of the walls of a nanopore.
Within the two-fluid model, the portion of the normal liquid dragged along with
the boundaries is dependent on the type of motion, and the resulting
anisotropic superfluid density plateaus far below unity at $T=0.5\kel$.  The
origin of the saturation is uncovered by computing the spatial distribution of
superfluidity, with only the core of the nanopore exhibiting any evidence of
phase coherence.  The superfluid core displays scaling behavior consistent with
Luttinger liquid theory, thereby providing an experimental test for the
emergence of a one dimensional quantum liquid.
\end{abstract}


\maketitle

\section{Introduction}
Superfluidity or dissipation-free flow,  is rooted in quantum mechanics
with the wave function of the entire fluid being described by an
emergent global macroscopic phase $\theta$.  In bulk ${}^4$He,  this breaking of
gauge symmetry  has dramatic consequences for the liquid below
the superfluid transition, $T_\lambda\simeq 2.17\kel$. It is well established that
superfluid helium can flow through extremely narrow constrictions, impenetrable
to the normal liquid, with a velocity $\vec{v}_s = (\hbar/m)\vec{\nabla}\theta$
limited only by a critical velocity first understood by Landau.  In ``rotating
bucket'' experiments, where a container of superfluid is rotated at an angular
frequency $\omega$, vortices can be spontaneously created, yielding a non-zero
quantum of fluid circulation $\kappa = \oint \vec{v}_s \cdot d\vec{r} =
\frac{h}{m} W$, where $W \in \mathbb{Z}$ is the topological winding number,
equal to the number of vortices within a closed circulation loop.  The
quantitative details of the superfluid state were first probed in the
celebrated Andronikashvili torsional oscillator experiment in 1946 where it was
determined that a \emph{superfluid fraction} of the total fluid does not
contribute to the classical moment of inertia. This observation led to the
development of Tisza's phenomenological two-fluid model where the superfluid
state is understood as two intertwined liquids, having normal ($\rho_{n}$) and
superfluid ($\rho_s$) components with total density $\rho = \rho_n + \rho_s$.

The superfluid-normal transition in bulk ${}^4$He is in the three dimensional
($3d$) XY universality class.  As the spatial dimension of the system is reduced,
the enhancement of fluctuations suppresses the transition temperature to zero
in $2d$ and precludes any long range ordered state in one dimension.  It is
intriguing to consider how the continuum two-fluid picture holds up in the low
dimensional limit where the bosonic helium system should be described by the
universal harmonic Luttinger liquid (LL) theory \cite{Haldane:1981gd} at low
energies.  Such a correlated liquid is strongly fluctuating with any phase
coherence decaying algebraically as a function of distance at $T =0\kel$.
Experimental realizations of low dimensional bosonic systems have been achieved
in ultra-cold atomic gases \cite{Cazalilla:2011dm} at low densities where the
interactions are expected to be weak and short ranged. At higher densities,
and in the presence of strong interactions and Galilean invariance, direct
observations of LL behavior are still lacking.  Previous experimental work in
low-$d$ quantum fluids has focused on superfluid helium confined to porous
materials with a radial length scale in the nanometer range, and there is
evidence for new quantum phases occurring at low temperature
\cite{Yamamoto:2008ed,Taniguchi:2011bx} where the dynamical response can be
understood in terms of LL theory \cite{Eggel:2011fj,Taniguchi:2013dk}.
Recently, experiments have demonstrated the feasibility of measuring the
superflow of helium through nanometer sized holes \cite{Savard:2011fe}. The
next step will be to systematically decrease the radius of the nanopore,
thereby providing a quasi-$1d$ flow geometry in which the superfluid properties
of helium can be measured.  Unequivocal evidence of LL behavior in ${}^4$He
filled nanopores will require a detailed understanding of the signatures of low
dimensional superfluidity in the crossover regime, where fluctuations and strong
interactions compete with the effects of confinement.  Towards this goal,
we have performed large scale numerical simulations measuring the superfluid
response of a strongly interacting confined quantum fluid of helium-4 at high
density. The results expose a breakdown of the two-fluid model of superfluidity
at the nanoscale and provide constraints on the experimental parameters needed
to observe an emergent $1d$ quantum liquid far from the previously observed
Tonks-Giradeau regime \cite{Kinoshita:2004jp, Paredes:2004fp}.

We begin by defining a model Hamiltonian that describes ${}^4$He confined
inside nanopores and provide some details of our quantum Monte Carlo (QMC)
method.  After a careful description of how superfluidity can be measured via
the linear response of the fluid to boundary motion, we present its temperature
dependence below the bulk superfluid transition.  An investigation of its
nanoscale properties identifies the presence of anisotropic superfluidity
originating from only the Luttinger liquid core region of pores with nanometer
radii.    

\section{Confined Helium-4}
The starting point is a system of helium-4 confined inside a nanopore of
radius $R$ and length $L$ formed as a cylindrical cavity in a slab of amorphous
silicon nitride.  The interactions between helium atoms, $U$ are modeled via the
Aziz potential \cite{Aziz:1979hs} while confinement is achieved by combining
the effects of short range repulsion with the walls of the pore and a long
range dispersion force between helium and the surrounding medium.
The result is a surface wetting potential, $V$,  with a deep attractive minimum
near the pore wall \cite{Tjatjopoulos:1988ec} where the Lennard-Jones
parameters have been chosen for Si$_3$N$_4$ ($\varepsilon = 10.22$~K, $\sigma=
2.628\ang$) to coincide with the nanofluidics experiments described above. 
The resulting quantum many-body Hamiltonian: 
\begin{equation}
\hat{H} = \sum_{i=1}^{N}\left[- \frac{\hbar^2}{2m}\hat{\vec{\nabla}}_i^2 +
\hat{V}(\vec{r_i}) \right] +
\sum_{i<j}\hat{U}(\vec{r}_i - \vec{r}_j)
\label{eqHamMicroscopic}
\end{equation}
where $N$ is the number of atoms with mass $m$, can be exactly simulated using
continuous space Worm Algorithm (WA) quantum Monte Carlo 
\cite{Boninsegni:2006gc}. Within the path integral formulation, this method
exploits the quantum-classical isomorphism, performing Metropolis sampling of
$(d+1)$-dimensional configurations of bosons that can be visualized as
worldlines or trajectories in an imaginary time ($\tau$) direction
\cite{Ceperley:1995gr}.  For helium at finite temperature, the worldlines obey
a periodicity condition in the additional dimension modulo identical particle
permutations, owing to their bosonic symmetry.  The superfluid response of the
low temperature system is directly linked to the existence and properties of
long connected worldline exchange cycles consisting of many individual atoms.

Our simulations employ a fixed chemical potential $\mu/k_\text{B}=-7.2\kel$ to
ensure helium atoms in pores of $L = 75\ang$ and $R=3.0-15.0\ang$ are in
thermal contact with a bath held at saturated vapor pressure for
temperatures between $0.5-2.25\kel$.  

\section{Superfluid Density}
\begin{figure}[t]
\begin{center}
\includegraphics[width=\columnwidth]{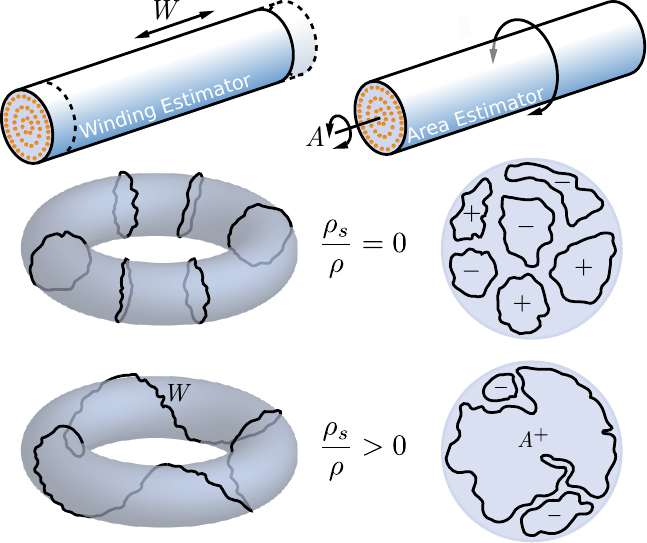}
\caption{\label{fig:sftypes} (color online). The origin of 
    superfluidity in ${}^4$He filled nanopores measured by the \emph{winding}
    (W, left column) and \emph{area} (A, right column) estimators in the path
    integral representation.  At high temperature, helium worldlines are short,
    containing only single particles and there is no superfluid fraction:
    $\rho_s/\rho = 0$. As the temperature is lowered, the worldlines may link
    with each other, winding around periodic boundary conditions in the
    simulation cell and having a large projected net area perpendicular to the
    axis of rotation.
}
\end{center}
\end{figure}

In QMC, the superfluid density is measured using linear response theory by
considering the effects of boundary motion \cite{Pollock:1987ta}.  Within the
two-fluid model, it is supposed that a superfluid fraction $\rho_s/\rho$ will
remain stationary while the normal portion $\rho_n/\rho$ will be dragged along
with the walls of the container.  In the nanopore geometry, two types of
motion, depicted in the first row of Fig.~\ref{fig:sftypes} are possible:
longitudinal motion of the walls along the cylindrical axis and a rotation
around it.  
%
%
These two types of response correspond to different measurements in the QMC,
related to the geometry and topology of particle worldlines, represented as
closed loops due to periodicity in imaginary time.   Within the naive two-fluid
picture, they should yield identical superfluid fractions and we can probe
this notion by concurrently measuring the dynamical response to both types of
boundary motion in the same nanopore.  

\subsection{Longitudinal Response}
In the case of longitudinal wall motion, (left column, Fig.~\ref{fig:sftypes}) we
consider a container with periodic boundary conditions along the axis of the
pore, and for the purposes of visualization, imagine the volume inside the pore
to be mapped onto the surface of a torus. The major circumference of the torus
is equal to $L$ while the minor one is $\hbar \beta$ with $\beta =
1/k_\mathrm{B}T$. In the path integral representation, the winding number is
given by: 
\begin{equation}
    W = \frac{1}{L}\sum_{i=1}^{N} \int_0^{\hbar \beta} d\tau\, 
    \left[ \frac{d z_i(\tau)}{d\tau} \right]
\end{equation}
where $z_i(\tau)$ is the $z$-component of $\vec{r}_i(\tau)$, the
$(d+1)$-dimensional position of particle $i$. It is equivalent to the number of
times the imaginary time trajectories of the $N$ particles wrap around the
periodic boundary conditions of the sample. In WA simulations, the imaginary
time $\tau$ must be discretized and we use $k_\text{B}\Delta \tau / \hbar =
0.004\kel^{-1}$ to minimize Trotter error. The
resulting superfluid density $\rhos{W}$ is related to the variance of the
distribution of winding numbers present\cite{Pollock:1987ta, Ceperley:1995gr}
through a \emph{winding estimator},
\begin{equation}
    \rhos{W} = \frac{m L}{\pi R^2 \hbar^2 \beta} \langle W^2
    \rangle
\label{eq:rhosW}
\end{equation}
where $m$ is the mass of a helium atom and $\langle \cdots \rangle$ indicates a
QMC average. At high temperature, the helium atoms behave classically, with
spatially localized worldlines containing only a single atom.  Long exchange
cycles are extremely unlikely and so $\langle W^2 \rangle = 0$.  As the
temperature is lowered, the kinetic energy can be reduced by linking worldlines
together.  Such particle exchanges can be efficiently sampled within the WA
using spatially local updates, producing configurations with extended
worldlines that wind around the periodic boundary conditions $(\langle W^2
\rangle \ne 0)$, producing a finite superfluid response.

\subsection{Rotational Response}
\begin{figure*}[t]
\begin{center}
\includegraphics[width=2\columnwidth]{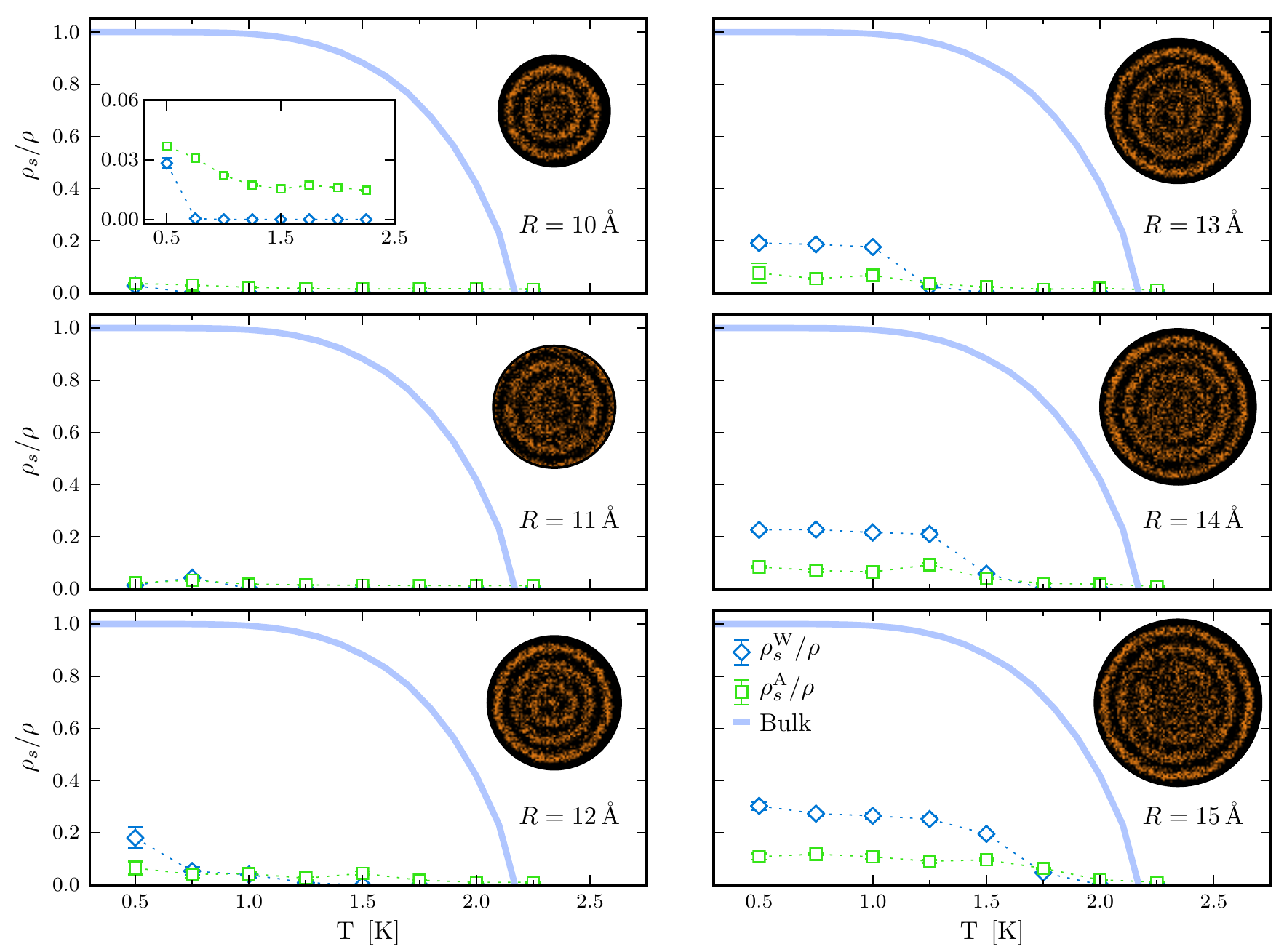}
\caption{\label{fig:sffull}(color online). A comparison of the superfluid
    fraction of helium confined inside nanopores with $L=75\ang$ (symbols)
    measured using the winding number and projected area of particle
    worldlines with the experimentally measured value taken from
    Ref.~[\onlinecite{Brooks:1977kz}] (line) for bulk ${}^4$He at saturated
    vapor pressure.  The upper right insets show instantaneous quantum Monte Carlo
    particle configurations projected onto the $z=0$ plane while the lower left
    inset in the first cell details superfluidity on a finer scale.  All panels
    share the legend shown in the lower right panel. 
}
\end{center}
\end{figure*}

An alternative approach, equivalent in the $d\ge3$ thermodynamic limit
\cite{Prokofev:2000ei}, measures the non-classical response of the fluid to a
small rotation.  The superfluid fraction $\rhos{A}/\rho$ is then equal to the
non-classical rotational moment of inertia fraction $(I_{\text{cl}} -
I)/I_{\text{cl}}$ where $I$ is the observed
moment of inertia and $I_{\text{cl}}$ is the total classical moment
of inertia.  The superfluid fraction defined in this way can be estimated in
the QMC by measuring the worldline area $A$ of closed particle trajectories
projected onto a plane perpendicular to the axis of rotation
\cite{Sindzingre:1989fo} through an \emph{area estimator},
\begin{equation} \rhos{A} = \frac{4 \rho
    m^2}{\hbar^2 \beta I_\text{cl}}
    \langle A^2 \rangle.
\label{eq:rhosA}
\end{equation}
For a rotation about the $z$-axis the path area is given by: 
\begin{equation}
    A = \frac{1}{2}\sum_{i=1}^{N} \int_0^{\hbar \beta}d \tau \left[\vec{r}_i(\tau)
\times \frac{d \vec{r}_i(\tau)}{d \tau}\right]_z.  
\end{equation}
At high temperature, the projected mean squared areas are uncorrelated and
$\sqrt{\langle A^2 \rangle} \sim \Lambda^2$ where $\Lambda = \sqrt{2 \pi \hbar^2
\beta / m}$ is the thermal de Broglie wavelength and $\rhos{A}/\rho \approx 0$
for large pores as seen in the right column of Fig.~\ref{fig:sftypes}.  As the
temperature is reduced and long exchange cycles become energetically favorable,
there is a distribution of finite projected areas and $\rhos{A}/\rho > 0$. The
caveat to this approach is that angular momentum conservation requires that the
dimensions of the simulation cell perpendicular to the axis of rotation do not
have periodic boundary conditions.  A quantum fluid confined inside a nanopore
thus provides an ideal geometry where we can directly compare the two
estimators of the superfluid fraction as shown in Fig.~\ref{fig:sffull}.  
%
%
The symbols correspond to QMC measurements performed using
Eqs.~(\ref{eq:rhosW}) and (\ref{eq:rhosA}) and the insets in the upper right
corners are instantaneous particle configurations projected on the plane $z=0$.
The solid line is the experimentally measured superfluid fraction of ${}^4$He at
saturated vapor pressure taken from Brooks and Donnelly \cite{Brooks:1977kz}
for comparison.  For pores with $R<9\ang$ we do not observe any superfluid
response above $T=0.5\kel$ while for $R\ge9\ang$, $\rhos{W,A}/\rho$ becomes
non-zero at a $R$-dependent onset temperature shifted below the bulk value of
$T_{\lambda}$.  This is the expected behavior for a quantum fluid constrained
inside a porous material where both $T_c$ and $\rho_s/\rho$ are reduced
\cite{Beamish:1983co,Finotello:1988ds} (see Ref.~[\onlinecite{Reppy:1992ch}]
for a review).

The observed superfluid response measured via the winding number is effectively
one dimensional, originating from flow along the pore axis, and any non-zero
value should be considered a finite size effect that will disappear as
$L\to\infty$. In this limit, superfluidity can arise from the dynamical
suppression of phase slips not captured in linear response theory
\cite{Eggel:2011fj}.  Another feature distinguishing the nanopore superfluid
fraction from that of bulk helium is an apparent saturation at low temperature
for both $\rhos{W}$ and $\rhos{A}$ at a value much less than one.  A hint at
the origin of this behavior can be observed by examining the spatial
configurations inside the pore in Fig.~\ref{fig:sffull}.  The interplay of
interactions between helium atoms as well as with the surrounding amorphous
Si$_3$N$_4$ leads to states exhibiting a series of cylindrical shells
\cite{Hernandez:2010dk, Rossi:2005hp, *Rossi:2006cj}, equivalent to the
formation of thin film layers of helium observed on $2d$ substrates including
silicate \cite{Boninsegni:2010fg}.

\subsection{Local Superfluidity}
\begin{figure}[t]
\begin{center}
\includegraphics[width=\columnwidth]{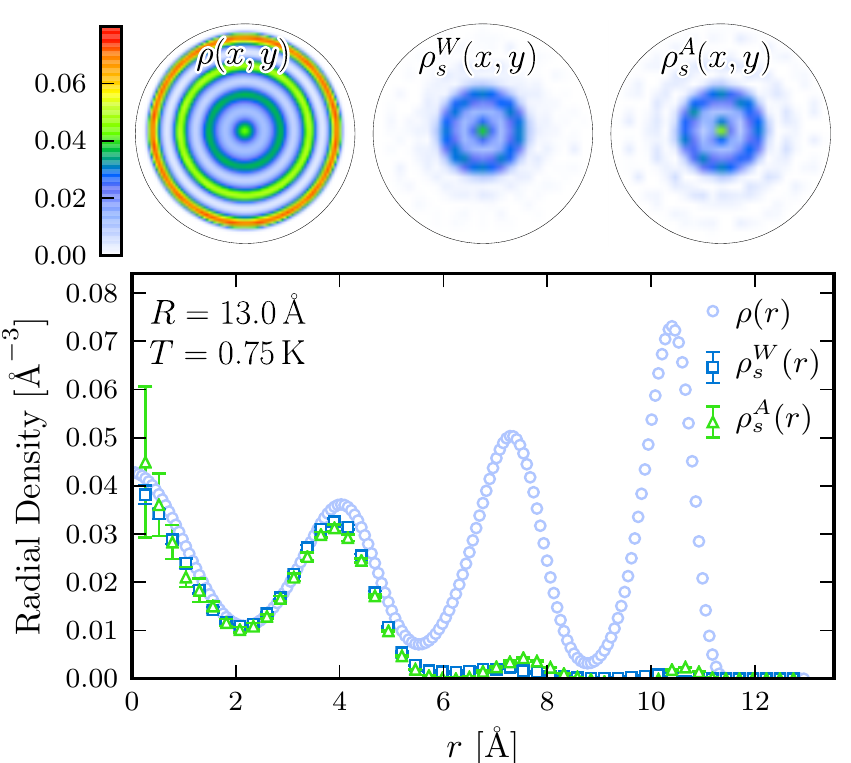}
\caption{\label{fig:locsf} (color online). The confined superfluid core seen by
    comparing the particle ($\rho$), winding
    ($\rhos{W}$) and area ($\rhos{A}$) superfluid number densities measured via
    quantum Monte Carlo simulations for a $R=13\ang$ radius nanopore of length
    $L=75\ang$ at $T=0.75\kel$. The first row shows the axially averaged local
    densities inside the pore while the lower plot contain an additional
    angular average.
}
\end{center}
\end{figure}

The competition between the tendency of bosons to delocalize at low $T$
and the strong geometrical confinement effects in the pore can be
investigated by measuring the local contribution of the superfluid density.
This was achieved by histogramming the radial $r$-dependence of the
winding number \cite{Khairallah:2005iz} or path area \cite{Kwon:2006en}
\begin{align}
\label{eq:rhosWlocal}
    \rhos{W}(r) &= \frac{m L^2}{\hbar^2 \beta} \langle W\cdot W(r) \rangle \\
\label{eq:rhosAlocal}
\rhos{A}(r) &= \frac{4m^2}{\hbar^2 \beta I_{\text{cl}}(r)} \langle A\cdot A(r) \rangle
\end{align}
with $W$ and $A$ the full pore values defined above while
\begin{equation}
    W(r) = \frac{1}{2 \pi r L^2} \sum_{i=1}^{N} \int_0^{\hbar \beta} d \tau
    \left[\frac{d z_i(\tau)}{d \tau}\right] \delta \left( r -
        r_i^\perp(\tau)\right)
\end{equation}
and 
\begin{equation}
    A(r) = \frac{1}{4 \pi r L} \sum_{i=1}^{N} \int_0^{\hbar \beta}\! d \tau
\left[\vec{r}_i(\tau) \times \frac{d \vec{r}_i(\tau)}{d \tau}\right]_z 
\delta \left( r - r_i^\perp(\tau)\right)
\end{equation}
where $I_{\text{cl}}(r) = m r^2$ is the classical moment inertia of a
single helium atom and $r_i^\perp(\tau) = \sqrt{x_i^2(\tau) +
y_i^2(\tau)}$.  A comparison of the average particle number density with the two
types of local superfluid density can be seen in Fig.~\ref{fig:locsf} for a
nanopore with $R=13\ang$ and $L=75\ang$ at $T=0.75\kel$.
%
%
The upper panels have been averaged over only the $z$-axis while the lower plot
contains the fully cylindrically symmetric radial values defined in
Eqs.~(\ref{eq:rhosWlocal}) and (\ref{eq:rhosAlocal}).  The saturation of the
total superfluid density seen in Fig~\ref{fig:sffull} can now be immediately
understood in terms of a spatial ``phase'' separation where only the inner
volume of the pore contains superfluid helium while the outer shells remain
nearly solid, adhering to the walls.  The results are qualitatively similar for
$R=12-15\ang$ with the two outermost shells making a negligible contribution to
the superfluid density. The local superfluid estimators are nearly
identical while their total values can differ by a factor of two.  This
presents no paradox due to their different normalizations: 
\begin{align}
    \frac{\rhos{W}}{\rho} &\equiv \frac{2\pi L}{N} \int_0^R r d r \rhos{W}(r), \\
    \frac{\rhos{A}}{\rho} &\equiv \frac{2\pi L m}{I_{\text{cl}}} 
    \int_0^R r d r [r^2 \rhos{A}(r)] 
\end{align}
which are required to account for local contributions to the classical moment of
inertia present in inhomogeneous fluids 
\begin{equation}
I_{\text{cl}}=2\pi L m \int_0^R r d r [r^2 \rho(r)] 
\end{equation} 
and provide consistency with linear response theory
\cite{Kwon:2006en}.  

\begin{figure}[t]
\begin{center}
\includegraphics[width=\columnwidth]{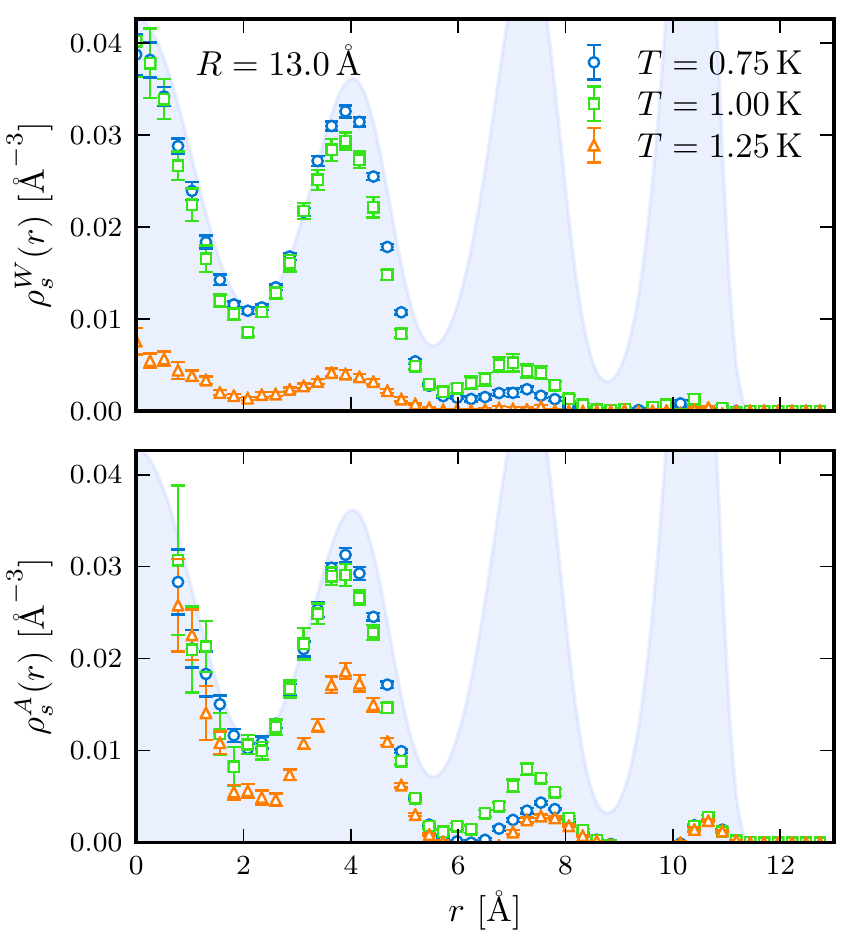}
\caption{\label{fig:locsfT} (color online).  The temperature dependence of the
    local superfluid density measured via the winding number (top) and area
    (bottom) estimator for a nanopore with radius $R=13.0\ang$ and length
    $L=75\ang$. The shaded region in the background corresponds to the particle
    density $\rho$ at $T=0.75\kel$.}
\end{center}
\end{figure}

The temperature dependence of the local superfluid density can also be studied,
and is shown for $R=13\ang$ in Figure~\ref{fig:locsfT}.
%
%
When the total superfluid density exhibits a plateau, there is only a minimal
dependence on temperature for the inner region of the pore.  As the temperature
is raised, the local superfluid density is reduced and begins to approach zero
between the shells indicating a suppression of inter-shell particle exchanges.
The non-zero and nearly temperature independent feature in $\rho_s^A(r)$
near the outer edge of the largest cylindrical shell appears for all radii
studied, and is responsible for the finite value of $\rho_s^A/\rho$ at high
temperature seen in the $R=10\ang$ inset of Fig.~\ref{fig:sffull}.  Its origin
is connected to the presence of enhanced correlations in the worldline area
when the radius of the cylindrical shell is on the order of the thermal de
Broglie wavelength.

\section{Luttinger Liquid Core}
\begin{figure}[t]
\begin{center}
\includegraphics[width=\columnwidth]{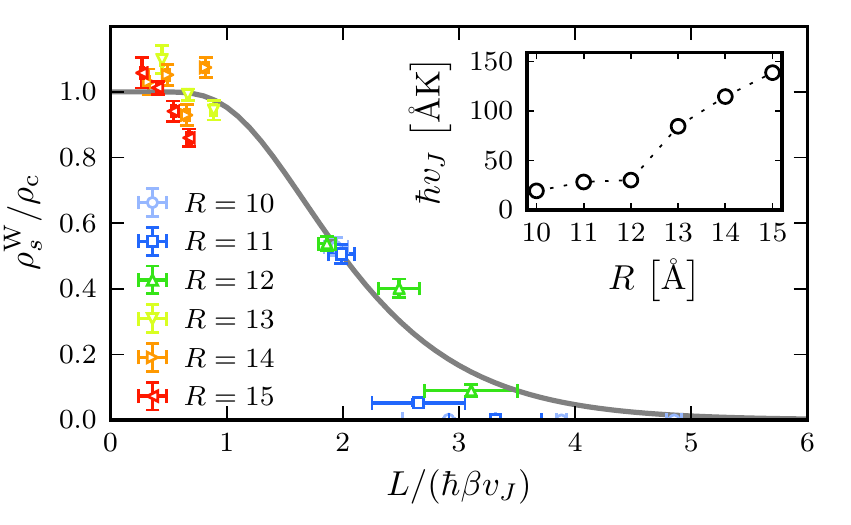}
\caption{\label{fig:sfll}(color online). The superfluid fraction of the core of
    nanopores for varying radii which can be collapsed onto the universal
    prediction from Luttinger liquid theory.  The inset shows the extracted
    value of the phase velocity $\hbar v_J$ obtained by fitting to the winding
    number estimator for each radius.
}
\end{center}
\end{figure}

The core region exhibiting a non-zero superfluid response is nearly one
dimensional, having a radius of $R \lesssim 6\ang$.  In $d=1$, fluctuations
preclude the existence of any long range superfluid order, and instead, the
helium system should be described at lowest order, by the linear quantum
hydrodynamics of Luttinger liquid theory \cite{Haldane:1981gd} with
effective Hamiltonian
\begin{equation}
H = \frac{\hbar v}{2\pi}\int_0^L d z \left[\frac{1}{K} \left(\partial_z \phi\right)^2 + K 
\left(\partial_z \theta\right)^2\right].
\label{eq:LLHam}
\end{equation}
The phases $\phi(z)$ and $\theta(z)$ are defined in terms of
the second quantized helium field operator $\psi^\dag(z) \sim \sqrt{\partial_z
\theta(z)}\mathrm{e}^{i\phi(z)}$ such that
$[\phi(z),\partial_{z^\prime}\theta(z')] = i \pi \delta(z-z^\prime)$.
Its low energy modes have dispersion $\varepsilon(k) =
\hbar v k$ and the value of the Luttinger parameter $K$ tunes the
system between algebraic superfluid ($K \ll 1$) or solid ($K\gg1$) order.
For a real physical system, the velocities $v_J \equiv v/K$ and $v_N \equiv v
K$ can be related to the parameters of the underling many-body
Hamiltonian.   By comparing the predictions of harmonic
LL theory, derived from the grand partition function $\mathcal{Z} =
\text{Tr}\,\exp[-\beta(H-\mu N)]$ with the measurements from finite temperature
QMC simulations, $v_J$ and $v_N$ can be determined. For quasi-$1d$ helium
confined inside nanopores with $R < 3\ang$, this has already been accomplished
\cite{DelMaestro:2011dh} but for larger radius pores, required the use of an
\emph{ad hoc} cutoff radius when analyzing QMC data.  The physical origin of
this cutoff is now fully understood as the radius of the superfluid core, and we
expect it to be described by LL theory\cite{DelMaestro:2010cz}:
\begin{equation}
    \frac{\rho_s^W}{\rho_c} = 1 - \frac{\pi \hbar \beta v_J}{L} 
    \left|\frac{\theta_3''\left(0,\mathrm{e}^{-2\pi \hbar \beta v_J/L}\right)}
    {\theta_3\left(0,\mathrm{e}^{-2\pi\hbar \beta v_J/L}\right)} \right|
\label{eq:rhosScale}
\end{equation}
where $\theta_3(z,q)$ is a Jacobi theta function with $\theta_3''(z,q) \equiv
\partial^2_z \theta_3(z,q)$ and $\rho_c = (N_c/N) \rho$ where $N_c$ is the
number of atoms in the core. For each radius, we have performed a
rescaling of the total superfluid response displayed in Fig.~\ref{fig:sffull}
and determined the velocity $v_J(R)$ through a fitting procedure that yields
the best collapse of all low temperature data onto Eq.~(\ref{eq:rhosScale}).
The results are displayed in Fig.~\ref{fig:sfll} where the temperature scaling
of the nanopore superfluidity is consistent with Luttinger liquid theory.
%
%

Much remains to be done, including confirming the predicted pore length
scaling of $\rhos{W}/\rho_c$ and evaluating the $R$-dependent LL parameter $K$.
In addition, it seems natural to contemplate the effects of disorder, surely
present in the pore walls, as well as the introduction of fermionic ${}^3$He
which may strongly alter superfluidity as bosonic exchanges will be suppressed
in $1d$.

\section{Conclusions}

We have performed large scale quantum Monte Carlo simulations for helium-4
confined inside short $75\ang$ pores with radii between $1-1.5~\mathrm{nm}$.  The
results show a finite and anisotropic superfluid response above $T=0.5~\kel$,
with a magnitude that is dependent on whether longitudinal or rotational motion
of the nanopore is considered.  The difference is large and arises from the
absence of any classical moment of inertia in the truly $1d$ limit where flow
is still possible.  Experiments probing this remarkable breakdown of the
two-fluid picture could be performed by comparing the superfluid fraction
measured by capillary flow and a nanoscale Andronikashvili torsional
oscillator.  Our results also indicate that when the radii of nanopores becomes
sufficiently small, the superfluid fraction may exhibit plateaus, increasing in
steps, due to the classical sticking of wetting layers near the pore walls.
This is in stark contrast to the usual smooth temperature dependence of
$\rho_s/\rho$ observed for bulk ${}^4$He and could provide a signature of the
crossover to $1d$ behavior. If the fraction of atoms adhering to the nanopores
walls could be discerned, possibly by comparing flow rates at high and low
temperature, an examination of the finite size and temperature scaling of the
superfluid density would confirm that confined low-dimensional helium is a
Luttinger liquid.  This would open up an exciting strongly interacting and high
density regime where the effective low energy theory can be experimentally
tested in systems with Galilean invariance. 

We acknowledge financial support from the University of Vermont,
NSERC (Canada), FQNRT (Qu\'ebec), and the Canadian Institute for Advanced
Research (CIFAR).  This research has been enabled by the use of computational
resources provided by the Vermont Advanced Computing Core supported by
NASA (NNX-08AO96G), WestGrid, SHARCNET and Compute/Calcul Canada.

\bibliographystyle{apsrev4-1.bst}
\bibliography{sfnanopores}

\begin{thebibliography}{28}%
\makeatletter
\providecommand \@ifxundefined [1]{%
 \@ifx{#1\undefined}
}%
\providecommand \@ifnum [1]{%
 \ifnum #1\expandafter \@firstoftwo
 \else \expandafter \@secondoftwo
 \fi
}%
\providecommand \@ifx [1]{%
 \ifx #1\expandafter \@firstoftwo
 \else \expandafter \@secondoftwo
 \fi
}%
\providecommand \natexlab [1]{#1}%
\providecommand \enquote  [1]{``#1''}%
\providecommand \bibnamefont  [1]{#1}%
\providecommand \bibfnamefont [1]{#1}%
\providecommand \citenamefont [1]{#1}%
\providecommand \href@noop [0]{\@secondoftwo}%
\providecommand \href [0]{\begingroup \@sanitize@url \@href}%
\providecommand \@href[1]{\@@startlink{#1}\@@href}%
\providecommand \@@href[1]{\endgroup#1\@@endlink}%
\providecommand \@sanitize@url [0]{\catcode `\\12\catcode `\$12\catcode
  `\&12\catcode `\#12\catcode `\^12\catcode `\_12\catcode `\%12\relax}%
\providecommand \@@startlink[1]{}%
\providecommand \@@endlink[0]{}%
\providecommand \url  [0]{\begingroup\@sanitize@url \@url }%
\providecommand \@url [1]{\endgroup\@href {#1}{\urlprefix }}%
\providecommand \urlprefix  [0]{URL }%
\providecommand \Eprint [0]{\href }%
\providecommand \doibase [0]{http://dx.doi.org/}%
\providecommand \selectlanguage [0]{\@gobble}%
\providecommand \bibinfo  [0]{\@secondoftwo}%
\providecommand \bibfield  [0]{\@secondoftwo}%
\providecommand \translation [1]{[#1]}%
\providecommand \BibitemOpen [0]{}%
\providecommand \bibitemStop [0]{}%
\providecommand \bibitemNoStop [0]{.\EOS\space}%
\providecommand \EOS [0]{\spacefactor3000\relax}%
\providecommand \BibitemShut  [1]{\csname bibitem#1\endcsname}%
\let\auto@bib@innerbib\@empty
\bibitem [{\citenamefont {Haldane}(1981)}]{Haldane:1981gd}%
  \BibitemOpen
  \bibfield  {author} {\bibinfo {author} {\bibfnamefont {F.~D.~M.}\
  \bibnamefont {Haldane}},\ }\href {\doibase 10.1103/PhysRevLett.47.1840}
  {\bibfield  {journal} {\bibinfo  {journal} {Phys. Rev. Lett.}\ }\textbf
  {\bibinfo {volume} {47}},\ \bibinfo {pages} {1840} (\bibinfo {year}
  {1981})}\BibitemShut {NoStop}%
\bibitem [{\citenamefont {Cazalilla}\ \emph {et~al.}(2011)\citenamefont
  {Cazalilla}, \citenamefont {Citro}, \citenamefont {Giamarchi}, \citenamefont
  {Orignac},\ and\ \citenamefont {Rigol}}]{Cazalilla:2011dm}%
  \BibitemOpen
  \bibfield  {author} {\bibinfo {author} {\bibfnamefont {M.}~\bibnamefont
  {Cazalilla}}, \bibinfo {author} {\bibfnamefont {R.}~\bibnamefont {Citro}},
  \bibinfo {author} {\bibfnamefont {T.}~\bibnamefont {Giamarchi}}, \bibinfo
  {author} {\bibfnamefont {E.}~\bibnamefont {Orignac}}, \ and\ \bibinfo
  {author} {\bibfnamefont {M.}~\bibnamefont {Rigol}},\ }\href
  {http://link.aps.org/doi/10.1103/RevModPhys.83.1405} {\bibfield  {journal}
  {\bibinfo  {journal} {Rev. Mod. Phys.}\ }\textbf {\bibinfo {volume} {83}},\
  \bibinfo {pages} {1405} (\bibinfo {year} {2011})}\BibitemShut {NoStop}%
\bibitem [{\citenamefont {Yamamoto}\ \emph {et~al.}(2008)\citenamefont
  {Yamamoto}, \citenamefont {Shibayama},\ and\ \citenamefont
  {Shirahama}}]{Yamamoto:2008ed}%
  \BibitemOpen
  \bibfield  {author} {\bibinfo {author} {\bibfnamefont {K.}~\bibnamefont
  {Yamamoto}}, \bibinfo {author} {\bibfnamefont {Y.}~\bibnamefont {Shibayama}},
  \ and\ \bibinfo {author} {\bibfnamefont {K.}~\bibnamefont {Shirahama}},\
  }\href {\doibase 10.1103/PhysRevLett.100.195301} {\bibfield  {journal}
  {\bibinfo  {journal} {Phys. Rev. Lett.}\ }\textbf {\bibinfo {volume} {100}},\
  \bibinfo {pages} {195301} (\bibinfo {year} {2008})}\BibitemShut {NoStop}%
\bibitem [{\citenamefont {Taniguchi}\ \emph {et~al.}(2011)\citenamefont
  {Taniguchi}, \citenamefont {Fujii},\ and\ \citenamefont
  {Suzuki}}]{Taniguchi:2011bx}%
  \BibitemOpen
  \bibfield  {author} {\bibinfo {author} {\bibfnamefont {J.}~\bibnamefont
  {Taniguchi}}, \bibinfo {author} {\bibfnamefont {R.}~\bibnamefont {Fujii}}, \
  and\ \bibinfo {author} {\bibfnamefont {M.}~\bibnamefont {Suzuki}},\ }\href
  {\doibase 10.1103/PhysRevB.84.134511} {\bibfield  {journal} {\bibinfo
  {journal} {Phys. Rev. B}\ }\textbf {\bibinfo {volume} {84}},\ \bibinfo
  {pages} {134511} (\bibinfo {year} {2011})}\BibitemShut {NoStop}%
\bibitem [{\citenamefont {Eggel}\ \emph {et~al.}(2011)\citenamefont {Eggel},
  \citenamefont {Cazalilla},\ and\ \citenamefont {Oshikawa}}]{Eggel:2011fj}%
  \BibitemOpen
  \bibfield  {author} {\bibinfo {author} {\bibfnamefont {T.}~\bibnamefont
  {Eggel}}, \bibinfo {author} {\bibfnamefont {M.~A.}\ \bibnamefont
  {Cazalilla}}, \ and\ \bibinfo {author} {\bibfnamefont {M.}~\bibnamefont
  {Oshikawa}},\ }\href {\doibase 10.1103/PhysRevLett.107.275302} {\bibfield
  {journal} {\bibinfo  {journal} {Phys. Rev. Lett.}\ }\textbf {\bibinfo
  {volume} {107}},\ \bibinfo {pages} {275302} (\bibinfo {year}
  {2011})}\BibitemShut {NoStop}%
\bibitem [{\citenamefont {Taniguchi}\ \emph {et~al.}(2013)\citenamefont
  {Taniguchi}, \citenamefont {Demura},\ and\ \citenamefont
  {Suzuki}}]{Taniguchi:2013dk}%
  \BibitemOpen
  \bibfield  {author} {\bibinfo {author} {\bibfnamefont {J.}~\bibnamefont
  {Taniguchi}}, \bibinfo {author} {\bibfnamefont {K.}~\bibnamefont {Demura}}, \
  and\ \bibinfo {author} {\bibfnamefont {M.}~\bibnamefont {Suzuki}},\
  }\href@noop {} {\bibfield  {journal} {\bibinfo  {journal} {Phys Rev B}\
  }\textbf {\bibinfo {volume} {88}},\ \bibinfo {pages} {014502} (\bibinfo
  {year} {2013})}\BibitemShut {NoStop}%
\bibitem [{\citenamefont {Savard}\ \emph {et~al.}(2011)\citenamefont {Savard},
  \citenamefont {Dauphinais},\ and\ \citenamefont {Gervais}}]{Savard:2011fe}%
  \BibitemOpen
  \bibfield  {author} {\bibinfo {author} {\bibfnamefont {M.}~\bibnamefont
  {Savard}}, \bibinfo {author} {\bibfnamefont {G.}~\bibnamefont {Dauphinais}},
  \ and\ \bibinfo {author} {\bibfnamefont {G.}~\bibnamefont {Gervais}},\ }\href
  {\doibase 10.1103/PhysRevLett.107.254501} {\bibfield  {journal} {\bibinfo
  {journal} {Phys. Rev. Lett.}\ }\textbf {\bibinfo {volume} {107}},\ \bibinfo
  {pages} {254501} (\bibinfo {year} {2011})}\BibitemShut {NoStop}%
\bibitem [{\citenamefont {Kinoshita}\ \emph {et~al.}(2004)\citenamefont
  {Kinoshita}, \citenamefont {Wenger},\ and\ \citenamefont
  {Weiss}}]{Kinoshita:2004jp}%
  \BibitemOpen
  \bibfield  {author} {\bibinfo {author} {\bibfnamefont {T.}~\bibnamefont
  {Kinoshita}}, \bibinfo {author} {\bibfnamefont {T.}~\bibnamefont {Wenger}}, \
  and\ \bibinfo {author} {\bibfnamefont {D.~S.}\ \bibnamefont {Weiss}},\
  }\href@noop {} {\bibfield  {journal} {\bibinfo  {journal} {Science}\ }\textbf
  {\bibinfo {volume} {305}},\ \bibinfo {pages} {1125} (\bibinfo {year}
  {2004})}\BibitemShut {NoStop}%
\bibitem [{\citenamefont {Paredes}\ \emph {et~al.}(2004)\citenamefont
  {Paredes}, \citenamefont {Widera}, \citenamefont {Murg}, \citenamefont
  {Mandel}, \citenamefont {F{\"o}lling}, \citenamefont {Cirac}, \citenamefont
  {Shlyapnikov}, \citenamefont {H{\"a}nsch},\ and\ \citenamefont
  {Bloch}}]{Paredes:2004fp}%
  \BibitemOpen
  \bibfield  {author} {\bibinfo {author} {\bibfnamefont {B.}~\bibnamefont
  {Paredes}}, \bibinfo {author} {\bibfnamefont {A.}~\bibnamefont {Widera}},
  \bibinfo {author} {\bibfnamefont {V.}~\bibnamefont {Murg}}, \bibinfo {author}
  {\bibfnamefont {O.}~\bibnamefont {Mandel}}, \bibinfo {author} {\bibfnamefont
  {S.}~\bibnamefont {F{\"o}lling}}, \bibinfo {author} {\bibfnamefont
  {I.}~\bibnamefont {Cirac}}, \bibinfo {author} {\bibfnamefont {G.~V.}\
  \bibnamefont {Shlyapnikov}}, \bibinfo {author} {\bibfnamefont {T.~W.}\
  \bibnamefont {H{\"a}nsch}}, \ and\ \bibinfo {author} {\bibfnamefont
  {I.}~\bibnamefont {Bloch}},\ }\href@noop {} {\bibfield  {journal} {\bibinfo
  {journal} {Nature}\ }\textbf {\bibinfo {volume} {429}},\ \bibinfo {pages}
  {277} (\bibinfo {year} {2004})}\BibitemShut {NoStop}%
\bibitem [{\citenamefont {Aziz}\ \emph {et~al.}(1979)\citenamefont {Aziz},
  \citenamefont {Nain}, \citenamefont {Carley}, \citenamefont {Taylor},\ and\
  \citenamefont {McConville}}]{Aziz:1979hs}%
  \BibitemOpen
  \bibfield  {author} {\bibinfo {author} {\bibfnamefont {R.~A.}\ \bibnamefont
  {Aziz}}, \bibinfo {author} {\bibfnamefont {V.~P.~S.}\ \bibnamefont {Nain}},
  \bibinfo {author} {\bibfnamefont {J.~S.}\ \bibnamefont {Carley}}, \bibinfo
  {author} {\bibfnamefont {W.~L.}\ \bibnamefont {Taylor}}, \ and\ \bibinfo
  {author} {\bibfnamefont {G.~T.}\ \bibnamefont {McConville}},\ }\href
  {\doibase doi:10.1063/1.438007} {\bibfield  {journal} {\bibinfo  {journal}
  {J. Chem. Phys.}\ }\textbf {\bibinfo {volume} {70}},\ \bibinfo {pages} {4330}
  (\bibinfo {year} {1979})}\BibitemShut {NoStop}%
\bibitem [{\citenamefont {Tjatjopoulos}\ \emph {et~al.}(1988)\citenamefont
  {Tjatjopoulos}, \citenamefont {Feke},\ and\ \citenamefont
  {Mann}}]{Tjatjopoulos:1988ec}%
  \BibitemOpen
  \bibfield  {author} {\bibinfo {author} {\bibfnamefont {G.~J.}\ \bibnamefont
  {Tjatjopoulos}}, \bibinfo {author} {\bibfnamefont {D.~L.}\ \bibnamefont
  {Feke}}, \ and\ \bibinfo {author} {\bibfnamefont {J.~A.}\ \bibnamefont
  {Mann}},\ }\href {\doibase doi: 10.1021/j100324a063} {\bibfield  {journal}
  {\bibinfo  {journal} {J. Phys. Chem.}\ }\textbf {\bibinfo {volume} {92}},\
  \bibinfo {pages} {4006} (\bibinfo {year} {1988})}\BibitemShut {NoStop}%
\bibitem [{\citenamefont {Boninsegni}\ \emph {et~al.}(2006)\citenamefont
  {Boninsegni}, \citenamefont {Prokof'ev},\ and\ \citenamefont
  {Svistunov}}]{Boninsegni:2006gc}%
  \BibitemOpen
  \bibfield  {author} {\bibinfo {author} {\bibfnamefont {M.}~\bibnamefont
  {Boninsegni}}, \bibinfo {author} {\bibfnamefont {N.~V.}\ \bibnamefont
  {Prokof'ev}}, \ and\ \bibinfo {author} {\bibfnamefont {B.~V.}\ \bibnamefont
  {Svistunov}},\ }\href {\doibase 10.1103/PhysRevE.74.036701} {\bibfield
  {journal} {\bibinfo  {journal} {Phys. Rev. E}\ }\textbf {\bibinfo {volume}
  {74}},\ \bibinfo {pages} {036701} (\bibinfo {year} {2006})}\BibitemShut
  {NoStop}%
\bibitem [{\citenamefont {Ceperley}(1995)}]{Ceperley:1995gr}%
  \BibitemOpen
  \bibfield  {author} {\bibinfo {author} {\bibfnamefont {D.~M.}\ \bibnamefont
  {Ceperley}},\ }\href {\doibase 10.1103/RevModPhys.67.279} {\bibfield
  {journal} {\bibinfo  {journal} {Rev. Mod. Phys.}\ }\textbf {\bibinfo {volume}
  {67}},\ \bibinfo {pages} {279} (\bibinfo {year} {1995})}\BibitemShut
  {NoStop}%
\bibitem [{\citenamefont {Pollock}\ and\ \citenamefont
  {Ceperley}(1987)}]{Pollock:1987ta}%
  \BibitemOpen
  \bibfield  {author} {\bibinfo {author} {\bibfnamefont {E.~L.}\ \bibnamefont
  {Pollock}}\ and\ \bibinfo {author} {\bibfnamefont {D.~M.}\ \bibnamefont
  {Ceperley}},\ }\href
  {http://links.isiglobalnet2.com/gateway/Gateway.cgi?GWVersion=2&SrcAuth=mekentosj&SrcApp=Papers&DestLinkType=FullRecord&DestApp=WOS&KeyUT=A1987L403500021}
  {\bibfield  {journal} {\bibinfo  {journal} {Phys. Rev. B}\ }\textbf {\bibinfo
  {volume} {36}},\ \bibinfo {pages} {8343} (\bibinfo {year}
  {1987})}\BibitemShut {NoStop}%
\bibitem [{\citenamefont {Prokof'ev}\ and\ \citenamefont
  {Svistunov}(2000)}]{Prokofev:2000ei}%
  \BibitemOpen
  \bibfield  {author} {\bibinfo {author} {\bibfnamefont {N.}~\bibnamefont
  {Prokof'ev}}\ and\ \bibinfo {author} {\bibfnamefont {B.}~\bibnamefont
  {Svistunov}},\ }\href {\doibase 10.1103/PhysRevB.61.11282} {\bibfield
  {journal} {\bibinfo  {journal} {Phys. Rev. B}\ }\textbf {\bibinfo {volume}
  {61}},\ \bibinfo {pages} {11282} (\bibinfo {year} {2000})}\BibitemShut
  {NoStop}%
\bibitem [{\citenamefont {Sindzingre}\ \emph {et~al.}(1989)\citenamefont
  {Sindzingre}, \citenamefont {Klein},\ and\ \citenamefont
  {Ceperley}}]{Sindzingre:1989fo}%
  \BibitemOpen
  \bibfield  {author} {\bibinfo {author} {\bibfnamefont {P.}~\bibnamefont
  {Sindzingre}}, \bibinfo {author} {\bibfnamefont {M.~L.}\ \bibnamefont
  {Klein}}, \ and\ \bibinfo {author} {\bibfnamefont {D.~M.}\ \bibnamefont
  {Ceperley}},\ }\href {\doibase 10.1103/PhysRevLett.63.1601} {\bibfield
  {journal} {\bibinfo  {journal} {Phys. Rev. Lett.}\ }\textbf {\bibinfo
  {volume} {63}},\ \bibinfo {pages} {1601} (\bibinfo {year}
  {1989})}\BibitemShut {NoStop}%
\bibitem [{\citenamefont {Brooks}\ and\ \citenamefont
  {Donnelly}(1977)}]{Brooks:1977kz}%
  \BibitemOpen
  \bibfield  {author} {\bibinfo {author} {\bibfnamefont {J.~S.}\ \bibnamefont
  {Brooks}}\ and\ \bibinfo {author} {\bibfnamefont {R.~J.}\ \bibnamefont
  {Donnelly}},\ }\href {\doibase 10.1063/1.555549} {\bibfield  {journal}
  {\bibinfo  {journal} {J. Phys. Chem. Ref. Data}\ }\textbf {\bibinfo {volume}
  {6}},\ \bibinfo {pages} {51} (\bibinfo {year} {1977})}\BibitemShut {NoStop}%
\bibitem [{\citenamefont {Beamish}\ \emph {et~al.}(1983)\citenamefont
  {Beamish}, \citenamefont {Hikata}, \citenamefont {Tell},\ and\ \citenamefont
  {Elbaum}}]{Beamish:1983co}%
  \BibitemOpen
  \bibfield  {author} {\bibinfo {author} {\bibfnamefont {J.}~\bibnamefont
  {Beamish}}, \bibinfo {author} {\bibfnamefont {A.}~\bibnamefont {Hikata}},
  \bibinfo {author} {\bibfnamefont {L.}~\bibnamefont {Tell}}, \ and\ \bibinfo
  {author} {\bibfnamefont {C.}~\bibnamefont {Elbaum}},\ }\href {\doibase
  10.1103/PhysRevLett.50.425} {\bibfield  {journal} {\bibinfo  {journal} {Phys.
  Rev. Lett.}\ }\textbf {\bibinfo {volume} {50}},\ \bibinfo {pages} {425}
  (\bibinfo {year} {1983})}\BibitemShut {NoStop}%
\bibitem [{\citenamefont {Finotello}\ \emph {et~al.}(1988)\citenamefont
  {Finotello}, \citenamefont {Gillis}, \citenamefont {Wong},\ and\
  \citenamefont {Chan}}]{Finotello:1988ds}%
  \BibitemOpen
  \bibfield  {author} {\bibinfo {author} {\bibfnamefont {D.}~\bibnamefont
  {Finotello}}, \bibinfo {author} {\bibfnamefont {K.~A.}\ \bibnamefont
  {Gillis}}, \bibinfo {author} {\bibfnamefont {A.}~\bibnamefont {Wong}}, \ and\
  \bibinfo {author} {\bibfnamefont {M.~W.~H.}\ \bibnamefont {Chan}},\
  }\href@noop {} {\bibfield  {journal} {\bibinfo  {journal} {Phys. Rev. Lett.}\
  }\textbf {\bibinfo {volume} {61}},\ \bibinfo {pages} {1954} (\bibinfo {year}
  {1988})}\BibitemShut {NoStop}%
\bibitem [{\citenamefont {Reppy}(1992)}]{Reppy:1992ch}%
  \BibitemOpen
  \bibfield  {author} {\bibinfo {author} {\bibfnamefont {J.~D.}\ \bibnamefont
  {Reppy}},\ }\href {\doibase 10.1007/BF00114905} {\bibfield  {journal}
  {\bibinfo  {journal} {J. Low Temp. Phys.}\ }\textbf {\bibinfo {volume}
  {87}},\ \bibinfo {pages} {205} (\bibinfo {year} {1992})}\BibitemShut
  {NoStop}%
\bibitem [{\citenamefont {Hern{\'a}ndez}(2010)}]{Hernandez:2010dk}%
  \BibitemOpen
  \bibfield  {author} {\bibinfo {author} {\bibfnamefont {E.~S.}\ \bibnamefont
  {Hern{\'a}ndez}},\ }\href {\doibase 10.1007/s10909-010-0238-8} {\bibfield
  {journal} {\bibinfo  {journal} {J. Low Temp. Phys.}\ }\textbf {\bibinfo
  {volume} {162}},\ \bibinfo {pages} {583} (\bibinfo {year}
  {2010})}\BibitemShut {NoStop}%
\bibitem [{\citenamefont {Rossi}\ \emph {et~al.}(2005)\citenamefont {Rossi},
  \citenamefont {Galli},\ and\ \citenamefont {Reatto}}]{Rossi:2005hp}%
  \BibitemOpen
  \bibfield  {author} {\bibinfo {author} {\bibfnamefont {M.}~\bibnamefont
  {Rossi}}, \bibinfo {author} {\bibfnamefont {D.~E.}\ \bibnamefont {Galli}}, \
  and\ \bibinfo {author} {\bibfnamefont {L.}~\bibnamefont {Reatto}},\ }\href
  {http://link.aps.org/doi/10.1103/PhysRevB.72.064516} {\bibfield  {journal}
  {\bibinfo  {journal} {Phys. Rev. B}\ }\textbf {\bibinfo {volume} {72}},\
  \bibinfo {pages} {064516} (\bibinfo {year} {2005})}\BibitemShut {NoStop}%
\bibitem [{\citenamefont {Rossi}\ \emph {et~al.}(2006)\citenamefont {Rossi},
  \citenamefont {Galli},\ and\ \citenamefont {Reatto}}]{Rossi:2006cj}%
  \BibitemOpen
  \bibfield  {author} {\bibinfo {author} {\bibfnamefont {M.}~\bibnamefont
  {Rossi}}, \bibinfo {author} {\bibfnamefont {D.~E.}\ \bibnamefont {Galli}}, \
  and\ \bibinfo {author} {\bibfnamefont {L.}~\bibnamefont {Reatto}},\ }\href
  {\doibase 10.1007/s10909-006-9265-x} {\bibfield  {journal} {\bibinfo
  {journal} {J. Low Temp. Phys.}\ }\textbf {\bibinfo {volume} {146}},\ \bibinfo
  {pages} {95} (\bibinfo {year} {2006})}\BibitemShut {NoStop}%
\bibitem [{\citenamefont {Boninsegni}(2010)}]{Boninsegni:2010fg}%
  \BibitemOpen
  \bibfield  {author} {\bibinfo {author} {\bibfnamefont {M.}~\bibnamefont
  {Boninsegni}},\ }\href
  {http://www.springerlink.com/index/10.1007/s10909-009-0143-1} {\bibfield
  {journal} {\bibinfo  {journal} {J. Low Temp. Phys.}\ }\textbf {\bibinfo
  {volume} {159}},\ \bibinfo {pages} {441} (\bibinfo {year}
  {2010})}\BibitemShut {NoStop}%
\bibitem [{\citenamefont {Khairallah}\ and\ \citenamefont
  {Ceperley}(2005)}]{Khairallah:2005iz}%
  \BibitemOpen
  \bibfield  {author} {\bibinfo {author} {\bibfnamefont {S.~A.}\ \bibnamefont
  {Khairallah}}\ and\ \bibinfo {author} {\bibfnamefont {D.~M.}\ \bibnamefont
  {Ceperley}},\ }\href {\doibase 10.1103/PhysRevLett.95.185301} {\bibfield
  {journal} {\bibinfo  {journal} {Phys. Rev. Lett.}\ }\textbf {\bibinfo
  {volume} {95}},\ \bibinfo {pages} {185301} (\bibinfo {year}
  {2005})}\BibitemShut {NoStop}%
\bibitem [{\citenamefont {Kwon}\ \emph {et~al.}(2006)\citenamefont {Kwon},
  \citenamefont {Paesani},\ and\ \citenamefont {Whaley}}]{Kwon:2006en}%
  \BibitemOpen
  \bibfield  {author} {\bibinfo {author} {\bibfnamefont {Y.}~\bibnamefont
  {Kwon}}, \bibinfo {author} {\bibfnamefont {F.}~\bibnamefont {Paesani}}, \
  and\ \bibinfo {author} {\bibfnamefont {K.~B.}\ \bibnamefont {Whaley}},\
  }\href {\doibase 10.1103/PhysRevB.74.174522} {\bibfield  {journal} {\bibinfo
  {journal} {Phys. Rev. B}\ }\textbf {\bibinfo {volume} {74}},\ \bibinfo
  {pages} {174522} (\bibinfo {year} {2006})}\BibitemShut {NoStop}%
\bibitem [{\citenamefont {Del~Maestro}\ \emph {et~al.}(2011)\citenamefont
  {Del~Maestro}, \citenamefont {Boninsegni},\ and\ \citenamefont
  {Affleck}}]{DelMaestro:2011dh}%
  \BibitemOpen
  \bibfield  {author} {\bibinfo {author} {\bibfnamefont {A.}~\bibnamefont
  {Del~Maestro}}, \bibinfo {author} {\bibfnamefont {M.}~\bibnamefont
  {Boninsegni}}, \ and\ \bibinfo {author} {\bibfnamefont {I.}~\bibnamefont
  {Affleck}},\ }\href {\doibase 10.1103/PhysRevLett.106.105303} {\bibfield
  {journal} {\bibinfo  {journal} {Phys. Rev. Lett.}\ }\textbf {\bibinfo
  {volume} {106}},\ \bibinfo {pages} {105303} (\bibinfo {year}
  {2011})}\BibitemShut {NoStop}%
\bibitem [{\citenamefont {Del~Maestro}\ and\ \citenamefont
  {Affleck}(2010)}]{DelMaestro:2010cz}%
  \BibitemOpen
  \bibfield  {author} {\bibinfo {author} {\bibfnamefont {A.}~\bibnamefont
  {Del~Maestro}}\ and\ \bibinfo {author} {\bibfnamefont {I.}~\bibnamefont
  {Affleck}},\ }\href {\doibase 10.1103/PhysRevB.82.060515} {\bibfield
  {journal} {\bibinfo  {journal} {Phys. Rev. B}\ }\textbf {\bibinfo {volume}
  {82}},\ \bibinfo {pages} {060515(R)} (\bibinfo {year} {2010})}\BibitemShut
  {NoStop}%
\end{thebibliography}%

\end{document}